\documentclass[twocolumn,pre,aps,showpacs]{revtex4}

\usepackage{graphicx}

\begin{document}


\title{
       On the intermittency exponent of the turbulent energy cascade
      }

\author{
        Jochen Cleve$^{1,2}$,
        Martin Greiner$^{3}$,
        Bruce R.\ Pearson$^{4}$, and
        Katepalli R.\ Sreenivasan$^{1}$
       }

\affiliation{$^{1}$International Centre for Theoretical Physics, Strada Costiera 11, 34014 Trieste, Italy;
         email: cleve@ictp.trieste.it, krs@ictp.trieste.it}
\affiliation{$^{2}$Institut f\"ur Theoretische Physik,
               Technische Universität,
               D-01062 Dresden, Germany}
\affiliation{$^{3}$Corporate Technology, Information {\&} Communications,
               Siemens AG, D-81730 M\"unchen, Germany;
         email: martin.greiner@siemens.com}
\affiliation{$^{4}$School of Mechanical, Materials,
               Manufacturing Engineering and Management,
               University of Nottingham,
               Nottingham NG7 2RD, United Kingdom;
         email: bruce.pearson@nottingham.ac.uk}

\date{\today}

\begin{abstract}
We consider the turbulent energy dissipation from one-dimensional
records in experiments using air and gaseous helium at cryogenic
temperatures, and obtain the intermittency exponent via the
two-point correlation function of the energy dissipation. The air
data are obtained in a number of flows in a wind tunnel and the
atmospheric boundary layer at a height of about 35 m above the
ground. The helium data correspond to the centerline of a jet
exhausting into a container. The air data on the intermittency
exponent are consistent with each other and with a trend that
increases with the Taylor microscale Reynolds number, $R_\lambda$,
of up to about 1000 and saturates thereafter. On the other hand,
the helium data cluster around a constant value at nearly all
$R_\lambda$, this being about half of the asymptotic value for the
air data. Some possible explanation is offered for this anomaly.
\end{abstract}

\pacs{47.27.Eq, 47.27.Jv, 47.53.+n, 05.40.-a, 02.50.Sk}



\maketitle

\section{Introduction}
That turbulent energy dissipation is intermittent in space has
been known since the seminal work of Batchelor \& Townsend
\cite{BT}. The characteristics of intermittency are best
expressed, at present, in terms of multifractals and the
multiplicity of scaling exponents; see e.g.\ Ref.~\cite{MEN91}. In
the hierarchy of the scaling exponents, the so-called
intermittency exponent characterizing the second-order behavior of
the energy dissipation, is the most basic. In this paper, we
address the Reynolds number variation of the intermittency
exponent and its asymptotic value (if one exists).

The intermittency exponent has been determined by a number of
authors in the past using several different methods; for a summary
as of some years ago, see Refs.\ \cite{SRE93,PRA97}. Recently,
Cleve, Greiner \& Sreenivasan \cite{CLE03a} evaluated these
methods and showed that the best procedure is to examine the
scaling of the two-point correlation function $\langle
\varepsilon(x+d) \varepsilon(x) \rangle$ of the energy dissipation
$\varepsilon$. Other procedures based on moments $\langle
\varepsilon_l^2 \rangle$ and $\langle\ln^2\varepsilon_l\rangle -
\langle\ln\varepsilon_l\rangle^2$ of the coarse-grained
dissipation $\varepsilon_l = (1/l) \int_l \varepsilon(x) \mathrm{d}x$, or
the power-spectrum $P(k) = | \varepsilon(k) |^2$, are corrupted by
the unavoidable surrogacy of the observed energy dissipation.

The follow-up effort \cite{CLE03b} was able to characterize and
understand the functional form of the two-point correlation
function beyond the power-law scaling range. Within the theory of
(binary) random multiplicative cascade processes, the finite-size
parametrization
\begin{equation}
\label{eq:one}
  \frac{ \left\langle
         \varepsilon(x+d) \varepsilon(x)
         \right\rangle }
       { \left\langle \varepsilon(x) \right\rangle^2 }
    =  c \left( \frac{L_\mathrm{casc}}{d} \right)^\mu
       + (1-c) \frac{d}{L_\mathrm{casc}}
\end{equation}
was derived, introducing the cascade length $L_\mathrm{casc}$ as a
meaningful physical upper length scale, this being similar to the
integral scale (as discussed later) but typically larger. Here,
$\mu$ is the intermittency exponent, and the first term is the
pure power-law part and the second term is the finite-size
correction. Comments on the constant $c$ will be given further
below. The one atmospheric and the three wind tunnel records
employed in Ref.\ \cite{CLE03b} were found to be in accordance
with this close-to-universal finite-size parametrization. Even for
flows at moderate Taylor-microscale Reynolds numbers $R_\lambda$,
Eq.\ (\ref{eq:one}) allowed an unambiguous extraction of $\mu$.
For the four data records, a weak dependence of $\mu$ on
$R_\lambda$ was observed in \cite{CLE03b} but was not commented
upon in any detail. The present analysis examines many more
records and attempts to put that Reynolds number dependence on
firmer footing.

\section{The data}
Two of the data records examined here come from the atmospheric
boundary layer (records a1 and a2) \cite{DHR00}, eight records
from a wind tunnel shear flow (records w1 to w8) \cite{PEA02} and
eleven records from a gaseous helium jet flow (records h1 to h11)
\cite{CHA00}. We find that all air experiments show an increasing
trend of the exponent towards 0.2 as the Reynolds number
increases. The helium data, on the other hand, show the
exceptional behavior of a Reynolds-number-independent ``constant"
of around 0.1. Some comments on this behavior are made.

\begin{figure}
\begin{centering}
\includegraphics[width=0.99\linewidth]{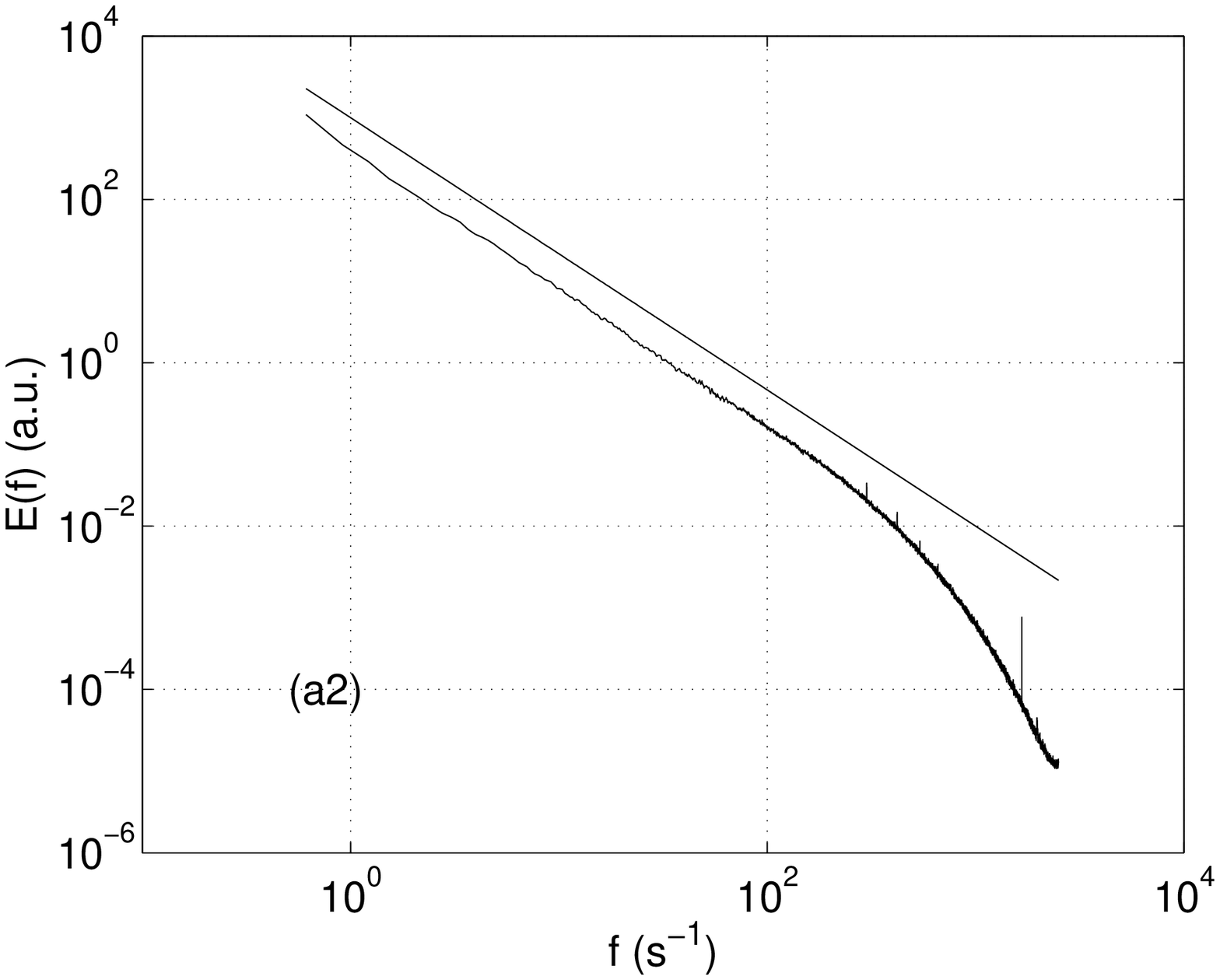}
\includegraphics[width=0.99\linewidth]{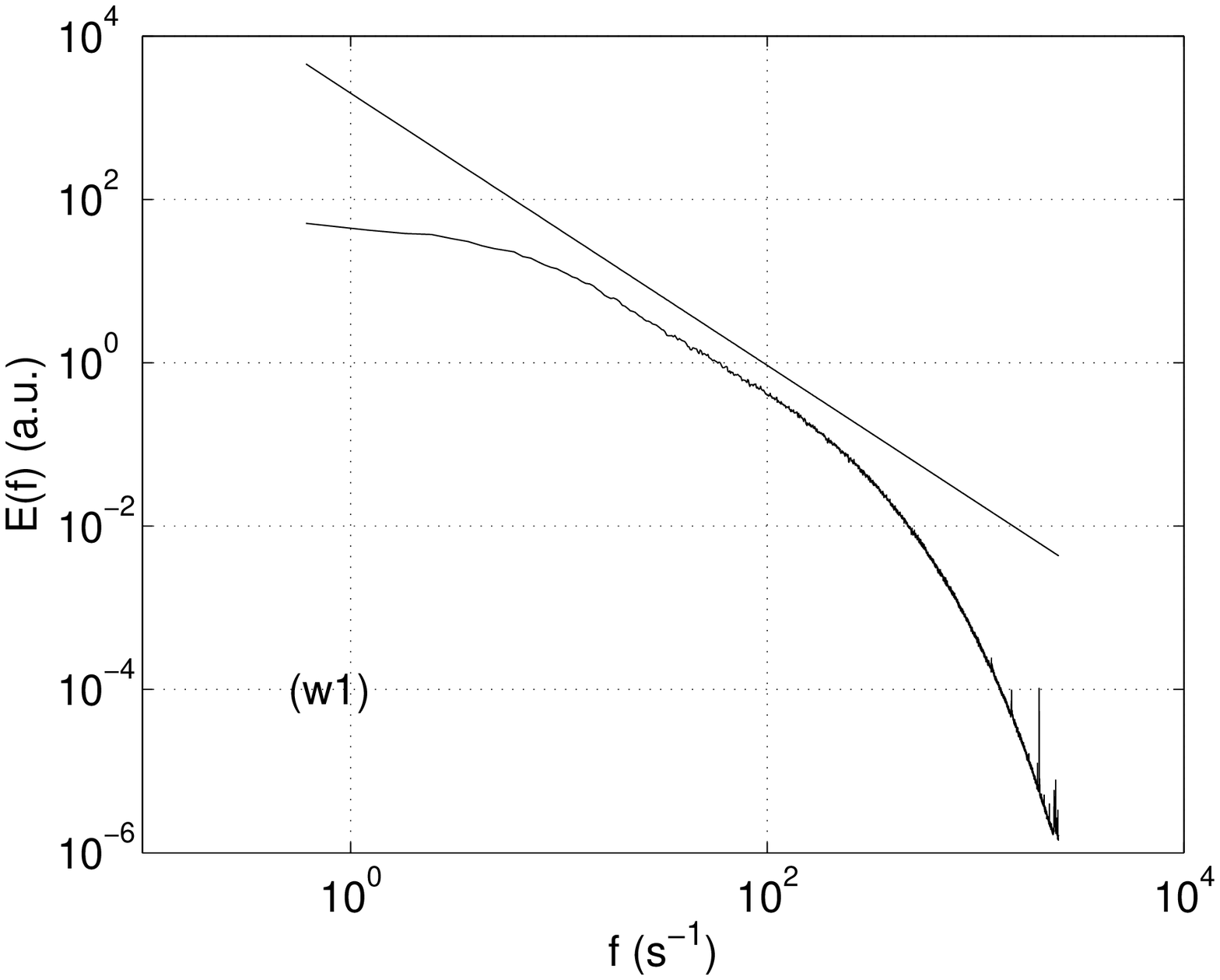}
\includegraphics[width=0.99\linewidth]{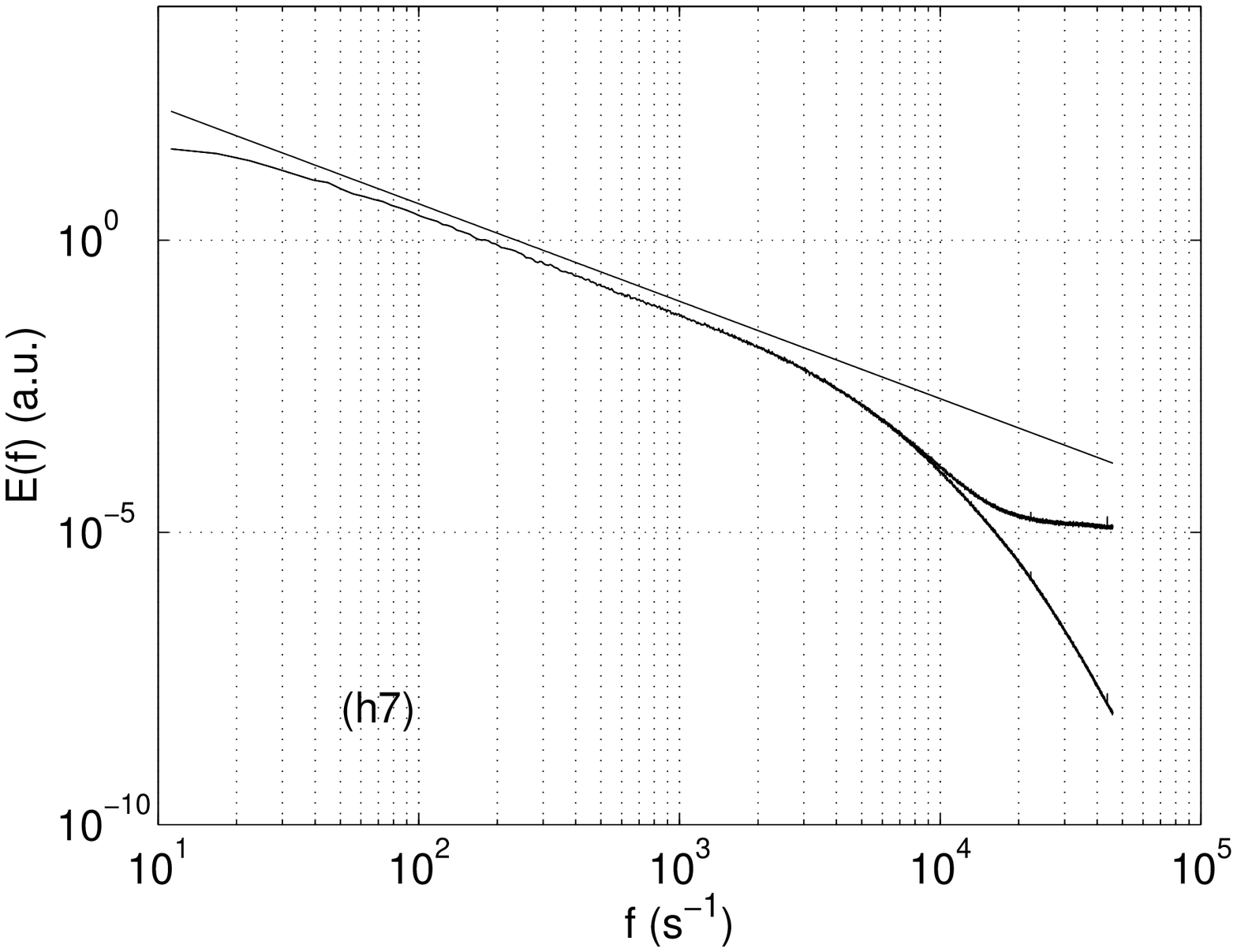}
\caption{Power spectral densities of the velocity fluctuations
for the records a2, w1 and h7. The sharp peaks are artifacts. For
the record h7, the filtered power spectral density is also shown.
} \label{fig:figure1}
\end{centering}
\end{figure}

It is useful to summarize the standard analysis in order to
emphasize the quality of the data. As is the standard practice,
energy dissipation is constructed via its one-dimensional
surrogate
\begin{equation}
\label{eq:two}
  \varepsilon(x)
    =  15 \nu \left(\frac{du}{dx}\right)^2
       \;,
\end{equation}
where the coordinate system is defined such that $u$ is the
component of the turbulent velocity pointing in the longitudinal
$x$ direction (the direction of mean motion). The coefficient
$\nu$ is the kinematic viscosity. Characteristic quantities such
as the Reynolds number $R_{\lambda} \equiv \sqrt{\langle
u^2\rangle} \lambda/\nu$, based on the Taylor microscale $\lambda
= \sqrt{\langle u^2 \rangle/\langle (\partial u/\partial x)^2
\rangle}$, the integral length $L$, the record length $L_\mathrm{record}$,
the resolution length $\Delta{x}$ and the hot-wire
length $l_w$ in units of the Kolmogorov dissipation scale
$\eta=(\nu^3/\langle\varepsilon\rangle)^{1/4}$ are listed in Table
\ref{tab:table1}.
\begin{table*}
 \caption{
  Taylor-microscale based Reynolds number $R_{\lambda}$,
  the integral length scale $L$ in units of the Kolmogorov scale $\eta$,
  the record length $L_\mathrm{record}$,
  the Taylor microscale $\lambda$,
  the resolution scale $\Delta x$
    ($=$ sampling time interval $\times$ mean velocity),
  the length of the hot wire $l_w$,
  the intermittency exponent $\mu$,
  the cascade length $L_\mathrm{casc}$ and
  the surrogacy cutoff length $\Lambda^*$ for the
  two atmospheric boundary layer data (a1,a2) \cite{DHR00},
  the eight wind tunnel data (w1--w8) \cite{PEA02} and
  the eleven sets from gaseous helium jet (h1--h11) \cite{CHA00} measurements.
  \label{tab:table1} }
 \begin{tabular}{l|cccccc|ccc}
   \hline
   data set & $R_{\lambda}$  & $L/\eta$         & $L_\mathrm{record}/L$
            & $\lambda/\eta$ & $\Delta x/\eta$  & $l_w/\eta$
            & $\mu$          & $L_\mathrm{casc}/\eta$ & $\Lambda^*/\eta$
   \\ \hline
   a1  & $9000$    & $5{\times}10^4$   & $1000$   & $187$ & $1.29$ & $1.755$
       & $0.216$   & $322743$          & $3.9$ \\
   a2  & $17000$   & $7.5{\times}10^4$ & $970$    & $246$ & $3.64$ & $1.534$
       & $0.202$   & $509354$          & $9.1$ \\
   w1  & $208$     & $539$             & $28000$  & $27$  & $2.42$ & $1.052$
       & $0.143$   & $1164$            & $24$  \\
   w2  & $306$     & $484$             & $102500$ & $35$  & $1.98$ & $1.780$
       & $0.155$   & $1939$            & $26$  \\
   w3  & $410$     & $697$             & $127700$ & $38$  & $2.67$ & $2.533$
       & $0.151$   & $2707$            & $27$  \\
   w4  & $493$     & $968$             & $193500$ & $44$  & $2.79$ & $3.382$
       & $0.145$   & $3228$            & $31$  \\
   w5  & $584$     & $1095$            & $88600$  & $44$  & $2.71$ & $0.890$
       & $0.172$   & $4062$            & $27$  \\
   w6  & $704$     & $1365$            & $117700$ & $48$  & $2.90$ & $1.079$
       & $0.173$   & $4343$            & $29$  \\
   w7  & $860$     & $1959$            & $89500$  & $53$  & $2.63$ & $1.580$
       & $0.176$   & $5513$            & $26$  \\
   w8  & $1045$    & $2564$            & $77500$  & $64$  & $2.97$ & $1.927$
       & $0.171$   & $7469$            & $27$  \\
   h1  & $85$      & $102$             & $197000$ & $22$  & $1.20$ & $0.040$
       & $0.12$    & $934$             & $10.8$\\
   h2  & $89$      & $101$             & $175000$ & $22$  & $1.05$ & $0.025$
       & $0.128$   & $472$             & $10.5$\\
   h3  & $124$     & $165$             & $100000$ & $26$  & $0.98$ & $0.068$
       & $0.102$   & $738$             & $17.7$\\%
   h4  & $208$     & $344$             & $85200$  & $33$  & $1.75$ & $0.088$
       & $0.154$   & $1258$            & $14.0$\\%
   h5  & $209$     & $277$             & $59000$  & $23$  & $0.97$ & $0.072$
       & $0.083$   & $1559$            & $10.7$\\
   h6  & $352$     & $606$             & $62400$  & $47$  & $2.25$ & $0.165$
       & $0.13$    & $2254$            & $22.5$\\%
   h7  & $463$     & $1011$            & $32100$  & $50$  & $1.93$ & $0.310$
       & $0.092$   & $10438$           & $25.1$\\
   h8  & $885$     & $1442$            & $40100$  & $47$  & $3.45$ & $0.763$
       & $0.089$   & $18954$           & $10.3$\\
   h9  & $929$     & $2064$            & $29800$  & $48$  & $3.67$ & $0.762$
       & $0.079$   & $8434$            & $18.3$\\%
   h10 & $985$     & $2144$            & $37800$  & $48$  & $4.83$ & $0.837$
       & $0.105$   & $23659$           & $14.5$\\
   h11 & $1181$    & $3106$            & $26900$  & $57$  & $4.97$ & $1.097$
       & $0.061$   & $14921$           & $19.9$\\
   \hline
 \end{tabular}
\end{table*}
To calculate the numerical value of $\lambda$ the method described
in \cite{ARO93} has been used. The integral length is defined as
the integral over the velocity autocorrelation function (using
Taylor's hypothesis). In the atmosphere, where the data do not
converge for very large values of the time lag, the
autocorrelation function is smoothly extrapolated to zero and the
integral is evaluated. This smoothing operation towards the tail
does not introduce measurable uncertainty in $L$. The energy
spectrum, which is illustrated in Fig.\ \ref{fig:figure1} for the
records a2, w1 and h7 as representatives of the three different
flow geometries with Reynolds numbers ranging from the small to
the very large side, follows an approximate $-5/3$-rds power over
the inertial range.
The wind tunnel records are relatively noise-free,
while the helium data are affected by instrumental noise
significantly, as evidenced by the flattening of the energy
spectrum for high wave numbers. The atmospheric data fall
somewhere in-between. The effect of this high-frequency noise, and
of removing it by a suitable filtering scheme, will be discussed
later.
\begin{figure}
\centering
\includegraphics[width=0.49\linewidth]{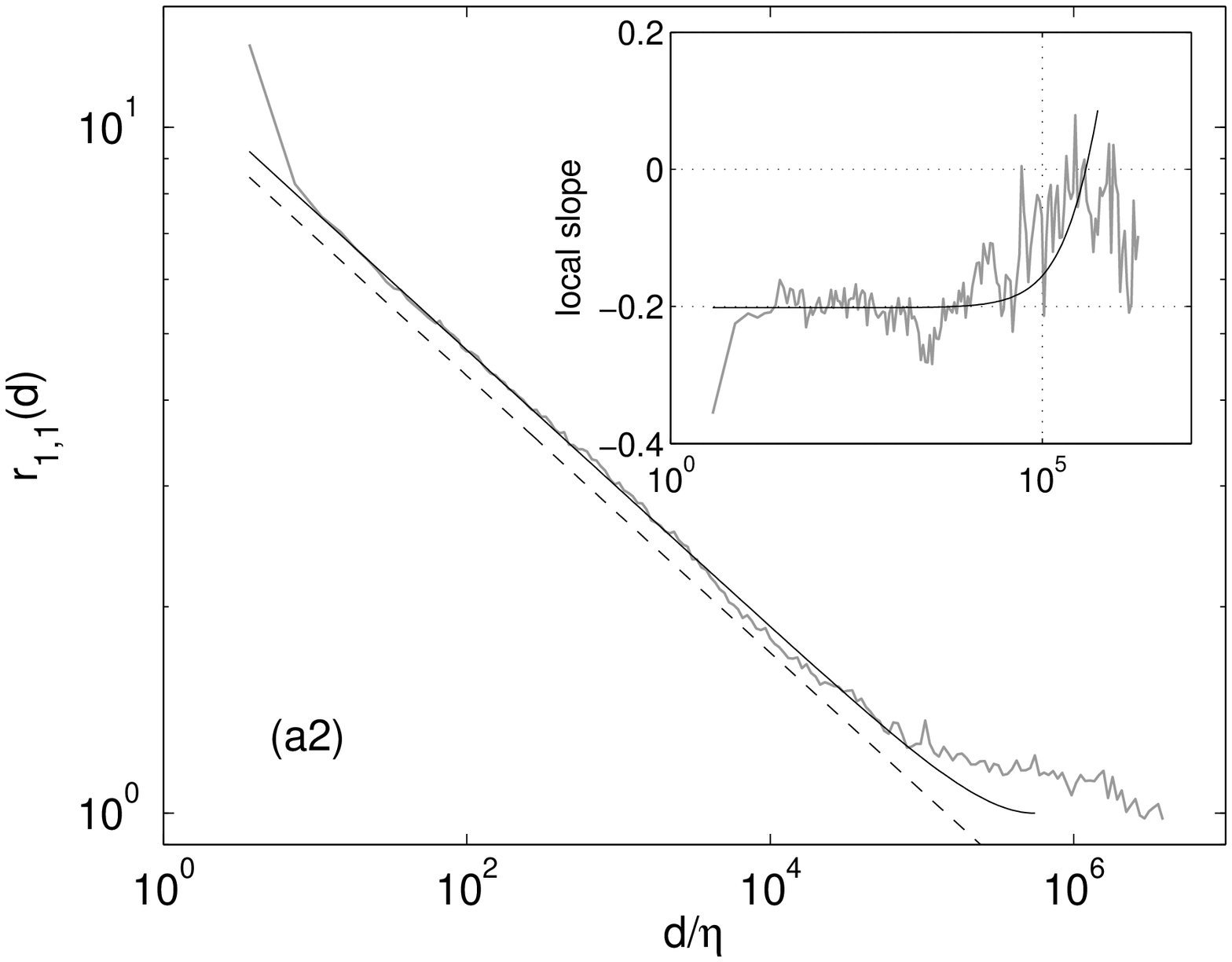}
\includegraphics[width=0.49\linewidth]{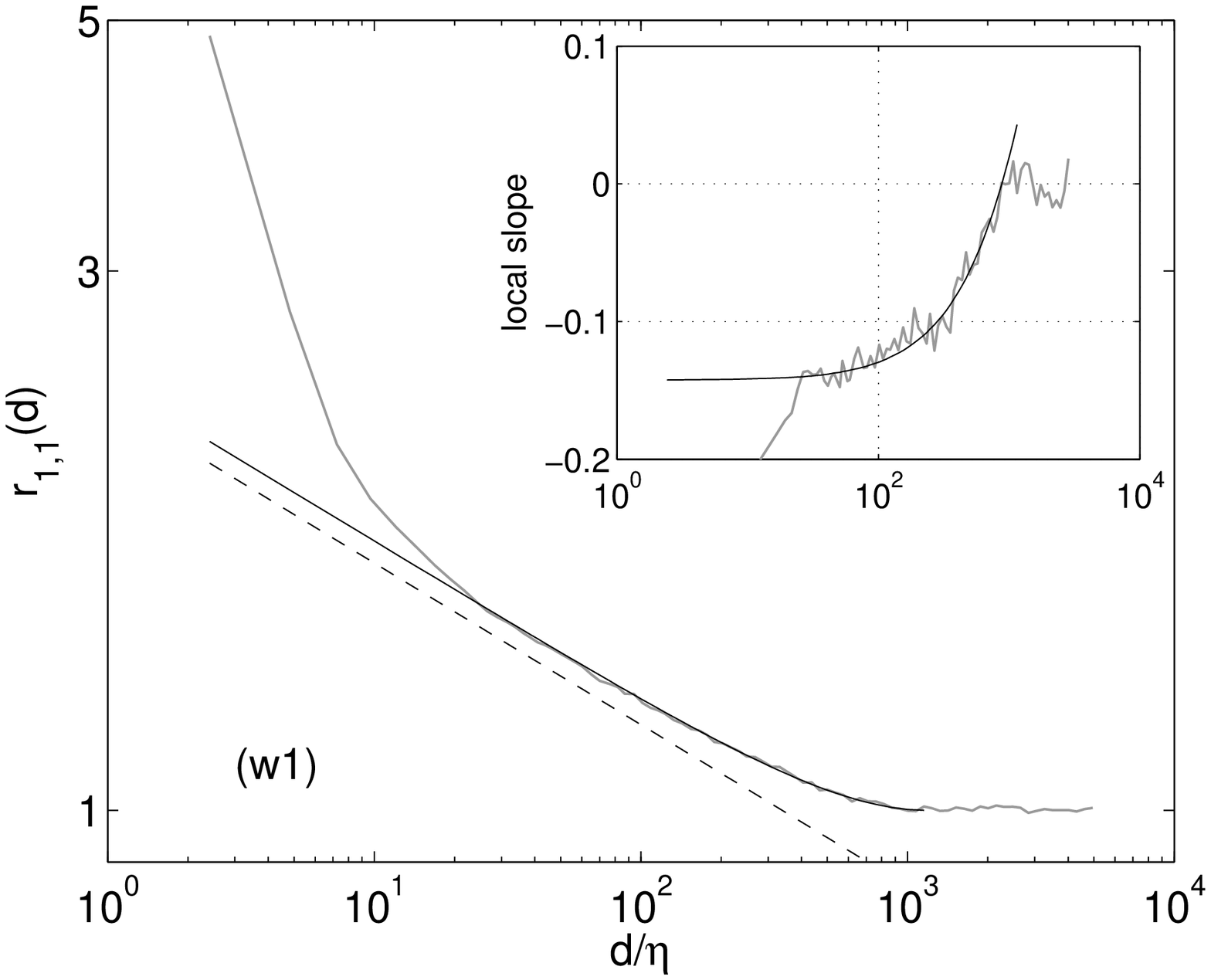}\\
\includegraphics[width=0.49\linewidth]{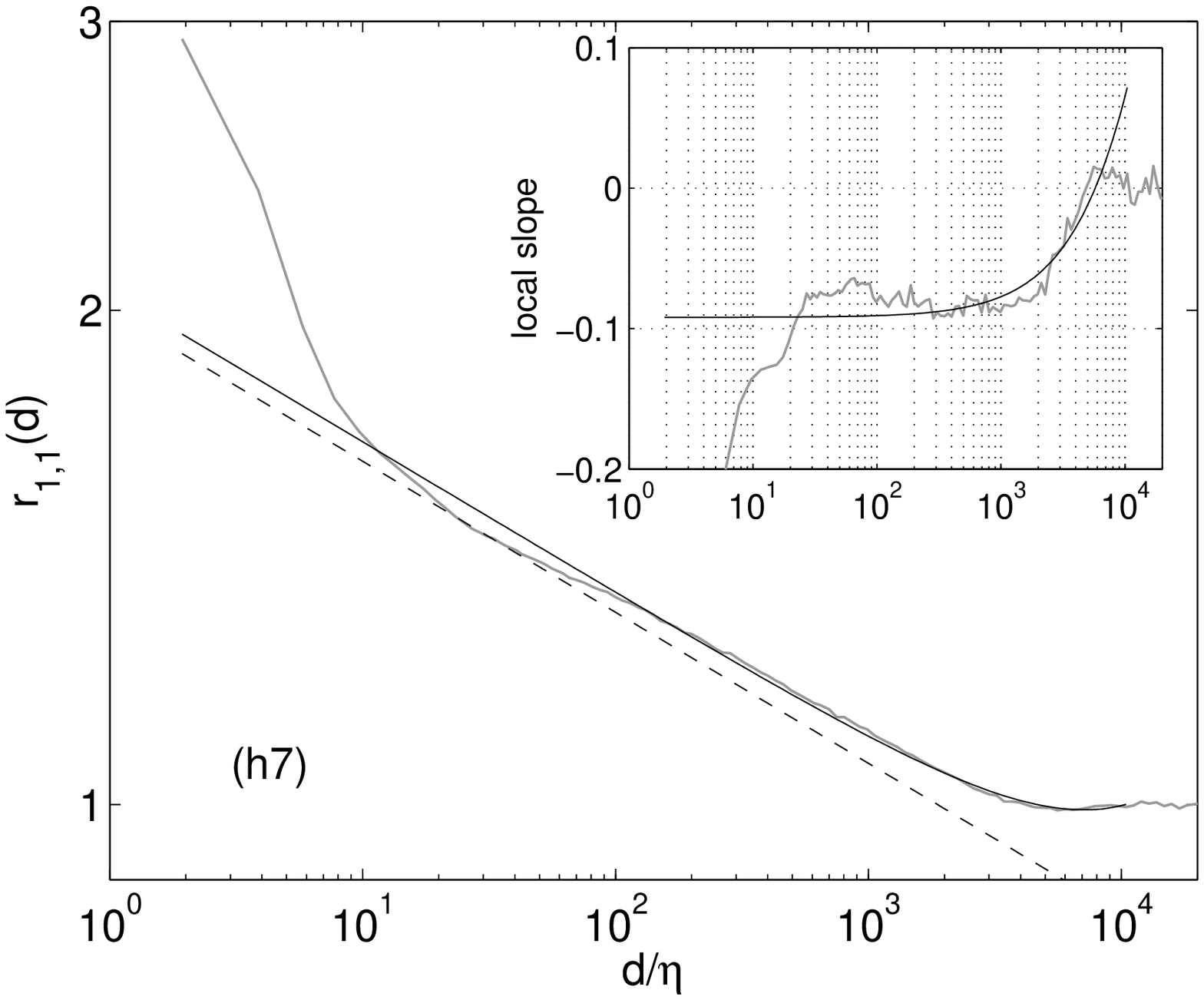}
\includegraphics[width=0.49\linewidth]{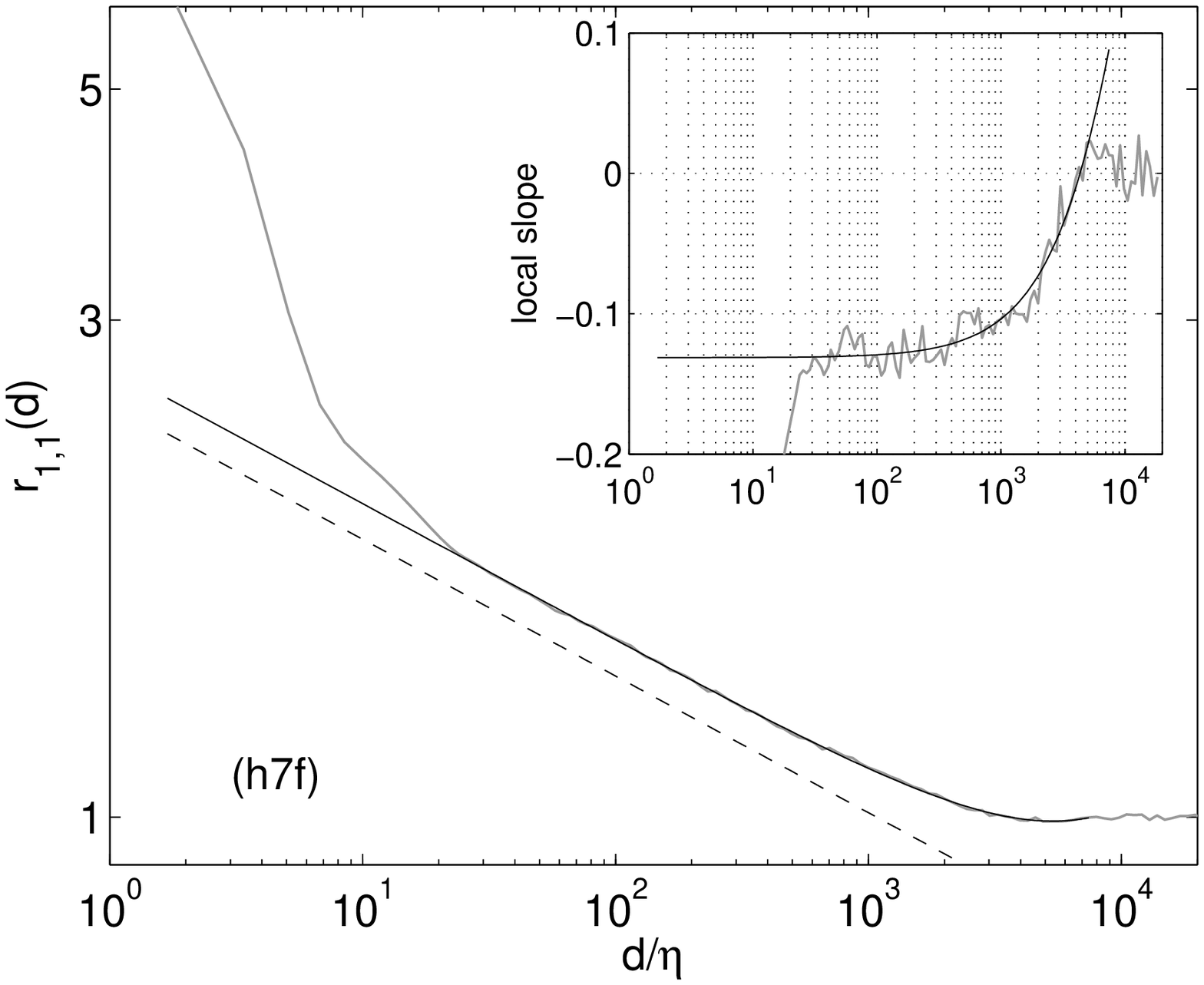}
\caption{Best fits of expression (\ref{eq:one}) to two-point
correlators extracted from data sets a2, w1, unfiltered h7 and
filtered h7 (h7f). The inset figures show the local slopes,
compared with the fits given by (1). For comparison, power-law
fits with the extracted intermittency exponent listed in Table I
(and $\mu=0.13$ for h7f) are shown as dashed straight lines drawn
with arbitrary shifts. } \label{fig:figure2}
\end{figure}
Again for the three representative cases a2, w1 and h7, Fig.\
\ref{fig:figure2} shows the two-point correlation defined in
(\ref{eq:one}) for the surrogate quantity (\ref{eq:two}) and
compares it to the best-fits for the proposed form of finite-size
parametrization given by Eq.\ (\ref{eq:one}). As is also the case
for all the other records, the agreement is quite substantial and
unambiguous. The upturn at small separation distances
$d<\Lambda^*$ has been explained in \cite{CLE03a} as the effect of
the surrogacy of the energy dissipation. Hence, we call
$\Lambda^*$ the surrogacy cutoff length. The two-point function
decorrelates at a length scale $L_\mathrm{casc}$ that is
substantially larger than the integral length scale. The cascade
length $L_\mathrm{casc}$ and the surrogacy cutoff length
$\Lambda^*$ are also listed in Table \ref{tab:table1} for all the
records inspected.

\section{Intermittency exponent}
To determine the intermittency exponent $\mu$, the data are
fitted to Eq.\ (\ref{eq:one}) using a best-fit algorithm (see
Fig.\ \ref{fig:figure2}). These values are also listed in Table
\ref{tab:table1}. The local slopes from the best fits, plotted as
insets in the figure, show that deviations from the pure power-law
become evident only towards large values of the separation
distance $d$. The dashed lines in each figure are pure power-laws
(with arbitrary shift) for the values of $\mu$ listed in the
Table.

\begin{figure}
\centering
\includegraphics[width=0.99\linewidth]{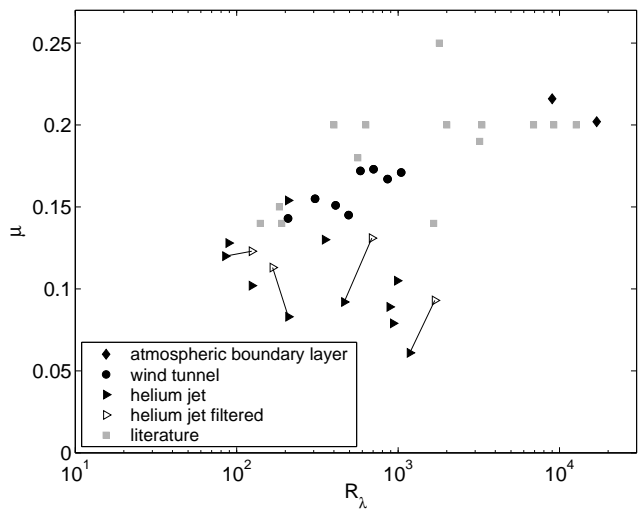}
\caption{The intermittency exponent, $\mu$, extracted from a best
fit of expression (\ref{eq:one}) to the two-point correlator of
the various data records, as a function of the Taylor-microscale
based Reynolds number. Also shown are some values quoted in the
literature. For some of the helium data, the lines show the shift
resulting from the application of the Wiener filter to remove
high-frequency noise.} \label{fig:figure3}
\end{figure}

For the atmospheric boundary layer, the analysis of two data sets
yields a value of about $0.2$ for the intermittency exponent $\mu$;
see Fig.\ \ref{fig:figure3}.
Note that, since the data set a1 contains both the longitudinal
and transverse velocity components, one can form different forms
of the surrogate energy dissipation. It was found in Ref.\
\cite{CLE03a} that all of them lead to the same value of the
intermittency exponent. The filtering of the data has no
measurable effect on the numerical value of $\mu$.

The wind-tunnel data w1-w8 seem to suggest a Reynolds number
dependence of the intermittency exponent for $R_\lambda$ of up to
about 1000. The value $\mu=0.2$ of the atmospheric boundary layer
is reached only for higher $R_\lambda$. Unfortunately there is no
laboratory data for $R_{\lambda} > 1000$ so that there is a gap
between the wind tunnel data and the atmospheric boundary layer
data. Nevertheless, all the air data taken together appear to be
consistent with a trend that increases with the Taylor microscale
Reynolds number up to an $R_\lambda$ of about 1000, and saturates
thereafter. This trend is also supported by results quoted in the
literature \cite{ANT81,ANT82,ANS84,KUZ92,SRE93,PRA97,MIJ01},
although the finite-size form (\ref{eq:one}) has not been employed
for the extraction of the intermittency exponent. The literature
values, shown in Fig.\ \ref{fig:figure3}, fill the gap between the
present wind tunnel and atmospheric data.

In contrast to the air data, the gaseous helium records h1-h11
show a different behavior (Fig.\ \ref{fig:figure3}). It appears
that, unlike the air data which show a gradual trend with
$R_\lambda$, leading to a saturation for $R_\lambda > 1000$, the
helium data yield an intermittency exponent that is flat with
$R_\lambda$ at a lower value of $0.1$. It remains an open question
as to why this is so. It would be important to settle this puzzle
and clarify if this special behavior has other consequences for
the helium jet data.

To make some progress, we examined the helium data more closely.
Perhaps the instrumental noise, which is seen in Fig.\
\ref{fig:figure1}c by the flattening of the energy spectrum for
high frequencies, affects the accuracy of the calculation of the
energy dissipation. To account for such effects, we applied a
Wiener filter to the data, see again Fig.\ \ref{fig:figure1}c, and
recomputed the two-point correlation; the result is shown in Fig.\
\ref{fig:figure2}d. The quality of the agreement with the
finite-size parametrization remained the same but the numerical
value for the intermittency exponent altered. Filtering produces
different amounts of shift for different sets of data; see again
Fig.\ \ref{fig:figure3}. The most extreme change of the numerical
value was found for h7, where the intermittency exponent changed
from $\mu=0.09$ in the unfiltered case to $\mu=0.13$ in the
filtered case. The difference between the two values can perhaps
be taken as the bounds for the error in the determination of the
intermittency exponent. Given this uncertainty, one cannot
attribute any trend with respect to the Reynolds number for the
helium data, and an average constant value of $\mu \approx 0.1$
seems to be a good estimate for all helium data.

Further questions relate to the spatial and temporal resolutions
of the hot wire. The temporal resolution in the helium case is
comparable to that in the air data (see Table \ref{tab:table1});
and, if anything, the ratio of the wire length to the smallest
flow scale, namely $l_w/\eta$, is better for helium experiments.
However, an important difference between the air data and the
helium data concerns the length to the diameter of the hot wire.
For air measurements, the ratio is usually of the order of a
hundred (about 140 for a1 and a2 and about 200 for w1 to w7),
while it is about 1/3 for h1-h11. In general, this is some cause
for concern because the conduction losses from the sensitive
element to the sides will be governed partly by this ratio, but
the precise effect depends on the conductivity of the material
with which the hot wires are made. For hot wires used in air
measurements, the material is a platinum-rhodium alloy, while for
those used in helium, the wire is made of Au-Ge sputtered on a
fiber glass. This issue has been discussed at some length for
similarly constructed hot wires of Emsellem et al.\
\cite{emsellem}. The conclusion there has not been definitive, but
the helium data discussed in \cite{emsellem} show another unusual
behavior: unlike the air data collected in \cite{sreeni}, the
flatness of the velocity derivative shows a non-monotonic behavior
with $R_\lambda$. See also figure 4 of Ref.\ \cite{gylf}. Whether
the two unusual behaviors of the helium data are related, and
whether they are in fact due to end losses, remains unclear and
cannot be confirmed without further study. A further comment is
offered towards the end of the paper.

The data on the surrogacy cutoff length $\Lambda^*$ does not show
a clear Reynolds number dependence. Referring again to Table
\ref{tab:table1}, it appears that $\Lambda^*$ is directly related
to neither $\lambda$ nor $\eta$. However, definitive statements
cannot be made because of the practical difficulty of locating
$\Lambda^*$ precisely.

\begin{figure}
\centering
\includegraphics[width=0.99\linewidth]{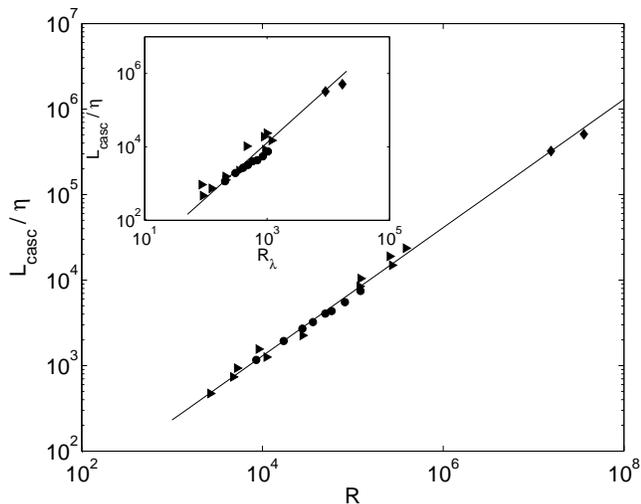}
\caption{The dependence on the Reynolds number of the ratio of the
cascade length $L_\mathrm{casc}$ to the Kolmogorov scale $\eta$.
The a-, w- and h-records are represented by diamonds, circles and
triangles, respectively. In the main graph, the straight line
indicates a power-law scaling with exponent $3/4$, and the
Reynolds number is defined as $R=u^{\prime}L_\mathrm{casc}/\nu$.
The inset shows the same data over $R_\lambda$ with the straight
line indicating the expected power-law scaling of $3/2$. The
prefactor for the main graph is $A=1.3$, for the inset $A=0.41$. }
\label{fig:figure4}
\end{figure}

Figure \ref{fig:figure4} illustrates the findings on the cascade
length ratio $L_\mathrm{casc}/\eta$.
The ratio increases with the large-scale Reynolds number $R =
u^{\prime}L_\mathrm{casc}/\nu$ as a power-law with the exponent of
$3/4$, exactly as anticipated if $L_\mathrm{casc}$ were
proportional to the integral scale. The ratio $L_\mathrm{casc}/L$
is not exactly a constant for all the data (as can be seen from
Table 1), but given the uncertainty in determining $L$ and the
absence of any systematic trend suggests that our supposition is
reasonable. This is further reinforced by the variation of
$L_\mathrm{casc}/\eta$ with respect to $R_\lambda$ (see inset),
which also follows the expected behavior. We should note that it
is difficult to single out the helium data in this respect.

\begin{figure}
\centering
\includegraphics[width=0.99\linewidth]{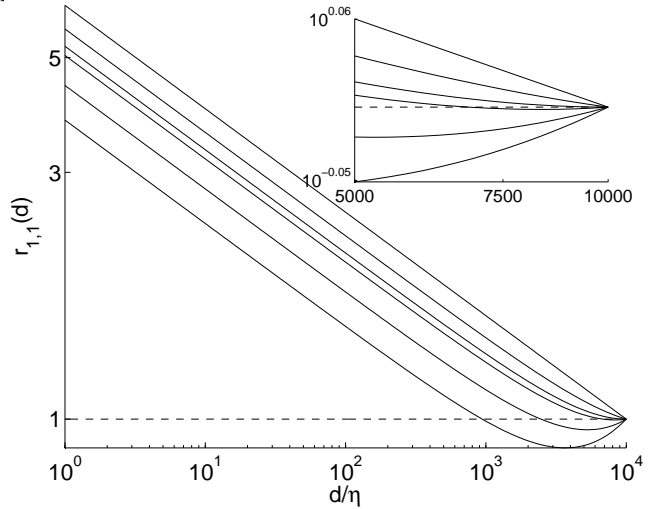}
\caption{
Two-point correlator (\ref{eq:one}) for various parameter values
$c = 0.6, 0.7, 0.8, 1/(1+\mu), 0.9, 1$
(from left to right). The other parameters are fixed to
$L_\mathrm{casc}/\eta=10^4$ and $\mu=0.20$.
}
\label{fig:figure5}
\end{figure}

There is more to learn from fitting the finite-size
parametrization (\ref{eq:one}) to the experimental two-point
correlations than merely extracting the intermittency exponent and
the cascade length. As revealed by a closer inspection of the
asymptotic behavior as $d \to L_\mathrm{casc}$, see again Fig.\
\ref{fig:figure2}, the two-point correlation of the atmospheric
boundary layer and wind tunnel records approach their asymptotic
value of unity from above, whereas for most of the gaseous helium
jet records the curve first swings a little below unity before
approaching the asymptotic value. The expression (\ref{eq:one}) is
flexible enough to reproduce even this behavior. The derivation of
(\ref{eq:one}) within the theory of binary random multiplicative
cascade processes, which has been presented in Ref.\
\cite{CLE03b}, also specifies the parameter
\begin{equation}
\label{eq:three}
  c =  \frac{\langle\Pi^2\rangle \langle q_Lq_R \rangle}
            {2 \langle q_{L/R}^2 \rangle - 1}
\end{equation}
in terms of cascade quantities. Normalized to
$\langle\Pi\rangle=1$, $\Pi$ represents the initial energy flux
density, which is fed into the cascade at the initial length scale
$L_\mathrm{casc}$. $\langle q_{L/R}^2 \rangle$ and $\langle q_Lq_R
\rangle$ are second-order moments of the bivariate probabilistic
cascade generator $p(q_L,q_R) = p(q_R,q_L)$, which we assume to be
symmetric. Again the normalization of the left and right random
multiplicative weights is such that $\langle q_{L/R} \rangle = 1$.
Note, that $\log_2 \langle q_{L/R}^2 \rangle = \mu$ is equal to
the intermittency exponent. Figure \ref{fig:figure5} shows various
graphs of the two-point correlation (\ref{eq:one}) with the
expression (\ref{eq:three}), where parameters $\mu$ and
$L_\mathrm{casc}$ have been kept fixed, but $c$ has been varied in the
range $0 < c < 1$.
We observe that for large $c$ the two-point correlation approaches
its asymptotic value from above, whereas for small $c$ it swings
below one before it reaches the asymptotic value from below. The
transition between these two behaviors occurs at $c \approx
\frac{1}{1+\mu}$. This translates to $\langle\Pi^2\rangle \langle
q_Lq_R \rangle = (2^{1+\mu}-1)/(1+\mu)$, which is $1.08$ for
$\mu=0.2$ and $1.04$ for $\mu=0.1$. Hence, we are tempted to
conclude that for the air data the fluctuation of the initial
energy flux density fed into the inertial-range cascade at the
upper length scale is somewhat larger than for the helium jet
data; this appears to be plausible and is one difference between
air and helium data. We also read this as an indication that
$\langle q_Lq_R \rangle < 1$, which is fulfilled if the left and
right multiplicative weight are anticorrelated to some extent. As
has already been discussed in a different context \cite{JOU00},
this anticorrelation is a clear signature that the
three-dimensional turbulent energy cascade conserves energy.

\section{Concluding remarks}
In summary, we state that the picture of the turbulent energy
cascade is robust and again confirmed by the excellent agreement
between the two-point correlation density predicted by random
multiplicative cascade theory and that extracted from various
experimental records. The cascade mechanism appears to be
universal, although its strength, as represented by the
intermittency exponent, seems to depend on the Reynolds number
except when it is very high. The discrepancy between the air data
on the one hand and the gaseous helium data on the other remains a
puzzle (despite some possible explanations offered), and is in
need of a fuller explanation.

\begin{acknowledgments}
The authors would like to thank Benoit Chabaud for providing his data.
\end{acknowledgments}


\end{document}